\documentclass[showpacs,preprintnumbers,two column]{revtex4-1}
\usepackage{amssymb}
\usepackage{amsmath}
\usepackage{graphicx}
\usepackage{dcolumn}
\usepackage{bm}

\begin{document}

\title{ Analytical solutions by squeezing to the anisotropic Rabi model in the nonperturbative deep-strong coupling regime }
\author{Yu-Yu Zhang, Xiang-You Chen}
\address{Department of Physics, Chongqing University, Chongqing
401330, People's Republic of China
}
\date{\today }

\begin{abstract}
A novel, unexplored nonperturbative deep-strong coupling (npDSC) achieved
in superconducting circuits has been studied in the anisotropic Rabi model
by the generalized squeezing rotating-wave approximation (GSRWA).
Energy levels are evaluated analytically from the
reformulated Hamiltonian and agree well with numerical
ones under a wide range of coupling strength.
Such improvement ascribes to deformation effects in the displaced-squeezed state
presented by the squeezed momentum variance, which are omitted
in the previous displaced state.
The population dynamics confirm the validity of our approach for the
npDSC strength. Our approach paves a way to the exploration of analysis in qubit-oscillator experiments
for the npDSC strength by the displaced-squeezed state.

\end{abstract}

\maketitle

\section{Introduction}

Quantum Rabi model~\cite{Rabi} describes the interaction of a two-level atom
with a single mode of the quantized electromagnetic field, which has been
completely solved by the rotating-wave approximation (RWA) on the assumption
of near resonance and weak coupling~\cite{jaynes}. Over recent decades,
progress has been made in increasing the strength of this interaction in
superconducting circuits~\cite{pforn,fumiki,Wallraff,Niemczyk,pfd,fedorov}.
Recent experimental progress has made it possible to
achieve a deep-strong coupling (DSC) strength
that approaches or exceeds the cavity frequency, $g/\omega\sim1$~\cite{pforn,fumiki}.
In this regime, the coupling is an order of
magnitude stronger than ultra-strong coupling (USC) strength previously
reported~\cite{Wallraff,Niemczyk,pfd,fedorov}, providing totally
different physics~\cite{casanova,de}. In the USC and DSC regimes,
the counter-rotating-wave (CRW) interaction are important and the RWA breaks down.
A generalization of the Rabi model
with independence coupling strengths of the rotating-wave and CRW
interactions, so-called the anisotropic Rabi model, has been attracting
interest~\cite{erlingsson,ye,xie,shen}.

Most studies describe the Rabi model involving the CRW terms
by different approximations in the USC regime due to the lack
of closed-form solutions~\cite{yu,zheng,irish,zhang1,agarwal,Ashhab,ying,plenio}. Since it is
understood physically that the atom-cavity interactions have two different
influences on the wave function of oscillators: displacement and deformation.
A generalized variational method (GVM) with variational
displacement~\cite{yu,zheng} improves the generalized RWA (GRWA)~\cite{irish,zhang1}
and adiabatic approximations~\cite{agarwal,Ashhab} with fixed
displacement in the USC regime, but is no longer valid for the DSC and high-frequency atom.
A perturbative treatment was reviewed when the atom part is a mere perturbation
for the DSC strength $g/\omega>1$, so-called as perturbative DSC~\cite{casanova,rossatto,alexandre}.
Between the USC and perturbative DSC regimes,
a novel, unexplored region is established as the npDSC regime~\cite{rossatto},
requiring an efficient, easy-to-implement analytical treatment.
As the coupling strength and atom frequency increase,
such approximations in the GVM and GRWA with only the displacement transformation is not
sufficient, and one need take account of the deformation of the oscillator state.
Recently, we have proposed the GSRWA with a displaced-squeezed state
to study the ground state of the Rabi model~\cite{zhang},
which improves the failure of the ground state obtained by the GVM
and GRWA for a wide range of coupling strengths. But an analytical
treatment for excited states remains elusive. Whether such substantial
improvement for the excited states in the npDSC regime remains
unexplored. So it is highly desirable
to give accuracy eigenstates and energies analytically in the npDSC regime
with the displaced-squeezed state, which includes both displacement and deformation effects.

The main purpose of this paper is to discuss excited states, deformation effects and
dynamics analytically by the GSRWA for the npDSC and high-frequency atom.
GSRWA combines the GVM with the
additional squeezing transformation and the standard RWA, resulting in a
more reasonable and closed-form solution.
The optimal displacement and squeezing parameters for excited states are expected to be
determined by eliminating the CRW terms and two-photon
process terms. Furthermore, we calculate the population dynamics to
compare the displaced-squeezed state and the displaced state to show which
is more stable in the npDSC regime.

The paper is outlined as follows: In Sec.~II, excited states and energies
are derived analytically using GSRWA for the anisotropic
Rabi model . Sec.~III is devoted to the suqeezing
effects by the quadrature variance for momentum operator.
In Sec.~IV, population dynamics of the atom
is discussed for a strong coupling strength. Finally, a brief
summary is given in Sec.~V.

\section{Anisotropic Rabi model}

The anisotropic Rabi Hamiltonian, describing a single cavity
mode coupled to a two-level atom, reads
\begin{equation}
H=\frac{1}{2}\Delta \sigma _{z}+a^{\dagger }a+g\left( a^{\dagger }\sigma
_{-}+a\sigma _{+}\right) +g\tau \left( a^{\dagger }\sigma _{+}+a\sigma
_{-}\right) ,  \label{Hamiltonian}
\end{equation}%
where $\Delta $ is atomic transition frequency, $g$ is the coupling
strength of rotating-wave interaction, $a^{\dagger }$ $\left(
a\right) $ is the photon creation (annihilation) operator of the single-mode
cavity with frequency $\omega $, and $\sigma _{k}(k=x,y,z)$ $\ $ are the
Pauli matrices. Here the relative weight between the rotating-wave
and CRW terms is adjusted by the parameter $\tau$. And the isotropic
Rabi model corresponds to $\tau =1$.

To facilitate the study, we write the Hamiltonian as
\begin{equation}
H=\frac{\Delta }{2}\sigma _{z}+\omega a^{\dagger }a+\alpha (a^{\dagger
}+a)\sigma _{x}+i\sigma _{y}\gamma \left( a^{\dagger }-a\right) ,
\label{horig}
\end{equation}%
with $\;\alpha =g\left(\tau+1 \right) /2$ and$\;\gamma =g\left( \tau
-1\right) /2$. Making use of a unitary transformation $U=\exp \left[ \beta
\sigma _{x}\left( a^{+}-a\right) \right] $ with the dimensionless
variational displacement $\beta $, we can obtain a transformed Hamiltonian $%
H_{1}=UHU^{\dagger }$,
\begin{eqnarray}
H_{1} &=&\omega a^{\dagger }a+\omega \beta ^{2}-2\beta \alpha +(\alpha
-\beta \omega )(a^{\dagger }+a)\sigma _{x}  \notag \\
&&+\frac{\Delta }{2}\{\sigma _{z}\cosh [2\beta \left( a^{\dagger }-a\right)
]-i\sigma _{y}\sinh [2\beta \left( a^{\dagger }-a\right) ]\}  \notag \\
&&+\gamma \left( a^{\dagger }-a\right) \{-\sigma _{z}\sinh [2\beta \left(
a^{\dagger }-a\right) ]  \notag \\
&&+i\sigma _{y}\cosh [2\beta \left( a^{\dagger }-a\right) ]\}.
\end{eqnarray}%
Such displacement transformation has been employed by the GVM and GRWA
for the isotropic Rabi model~\cite{irish,zhang1,yu},
which considers the displacement of the oscillator state and omit
the deformations effects induced by the coupling between the oscillator and atom. It is absolutely nontrivial
to extend the treatment to the anisotropic Rabi model, and to employ
an additional unitary transformation
\begin{equation}
S=e^{\lambda (a^{2}-a^{+2})}
\end{equation}%
with the dimensionless variational squeezing $\lambda $, which yields $%
SaS^{+}=a^{\dagger }\sinh 2\lambda +a\cosh 2\lambda $ and $Sa^{\dagger
}S^{+}=a\sinh 2\lambda +a^{\dagger }\cosh 2\lambda $. Then the Hamiltonian $%
H_{2}=SH_{1}S^{+}=H_{0}^{\prime }+H_{1}^{\prime }$ takes the form
\begin{eqnarray}
H_{0}^{\prime } &=&\eta _{0}+\eta _{1}\omega a^{\dagger }a+\sigma _{z}\{%
\frac{\Delta }{2}\cosh [2\beta \eta \left( a^{\dagger }-a\right) ]  \notag
\label{h0} \\
&&-\gamma \left( a^{\dagger }-a\right) \eta \sinh [2\beta \eta \left(
a^{\dagger }-a\right) ]\}+\eta _{2}(a^{\dagger 2}+a^{2}), \\
H_{1}^{\prime } &=&\eta _{3}\sigma _{x}(\alpha -\omega \beta )(a^{\dagger
}+a)+i\sigma _{y}\{-\frac{\Delta }{2}\sinh [2\beta \eta \left( a^{\dagger
}-a\right) ]  \notag \\
&&+\gamma \left( a^{\dagger }-a\right) \eta \cosh [2\beta \eta \left(
a^{\dagger }-a\right) ]\}.
\end{eqnarray}%
where $\eta _{0}=\omega\sinh ^{2}2\lambda +\omega\beta ^{2}-2\beta \alpha $, $\eta
_{1}=(\cosh ^{2}2\lambda +\sinh ^{2}2\lambda )$, $\eta _{2}=\cosh 2\lambda
\sinh 2\lambda $, $\eta _{3}=(\cosh 2\lambda +\sinh 2\lambda )$ and $\eta
=\cosh 2\lambda -\sinh 2\lambda $.

The additional squeezing transformation captures effects of the deformations
of the oscillator state, providing
a displaced-squeezed oscillator state instead of the previous displaced state.
On the other hand, the squeezing transformation introduces
the two-excitation terms $a^{\dagger 2}$ and $a^{2}$,
which is accounted for the two-photon process. In contrast to the GVM with only
the displacement transformation, it is expected to exhibits a substantial improvements of our approach.

Since $\cosh \left[ 2\beta \eta \left( a^{\dagger }-a\right) \right] $ and $%
\sinh \left[ 2\beta \eta \left( a^{\dagger }-a\right) \right] $ are the even
and odd functions, we can expand the functions by keeping leading terms as,
\begin{eqnarray}
\cosh \left[ 2\beta \eta \left( a^{\dagger }-a\right) \right] &=&G^{0}\left(
a^{\dagger }a\right) +G^{2}\left( a^{\dagger }a\right) a^{\dagger 2}  \notag
\\
&&+a^{2}G^{2}\left( a^{\dagger }a\right) +O(\beta ^{4}\eta ^{4}),
\end{eqnarray}%
\begin{equation}
\sinh \left[ 2\beta \eta \left( a^{\dagger }-a\right) \right] =F\left(
a^{\dagger }a\right) a^{\dagger }-aF\left( a^{\dagger }a\right) +O(\beta
^{3}\eta ^{3}),
\end{equation}%
where $G^{0}\left( a^{\dagger }a\right) $, $G^{2}\left( a^{\dagger }a\right)
$ and $F\left( a^{\dagger }a\right) $ $(i=0,1,2,...)$ are the coefficients
dependent on the oscillator number operator $a^{\dagger }a$. In the
oscillator basis $|n\rangle $, the coefficient $G^{0}\left( a^{\dagger
}a\right) $ can be expressed explicitly as
\begin{equation}
G_{n,n}^{0}=\langle n|\cosh [2\beta \eta (a^{\dagger }-a)]|n\rangle
=e^{-2\beta ^{2}\eta ^{2}}L_{n}(4\beta ^{2}\eta ^{2}),  \notag
\end{equation}%
with the Laguerre polynomials $L_{n}^{m-n}(x)$. And the coefficient $%
G^{2}\left( a^{\dagger }a\right)$ corresponding to two-excitation terms is
derived as $G_{n+2,n}^{2}$ in the Appendix A. Since the terms $F\left(
a^{\dagger }a\right) a^{\dagger }$ and $aF\left( a^{\dagger }a\right) $
involve creating and eliminating a single photon, the coefficient $%
F(a^{\dagger }a)$ of one-excitation terms is derived as
\begin{eqnarray}
F_{n+1,n} &=&\frac{1}{\sqrt{n+1}}\left\langle n+1\right\vert \sinh \left[
2\alpha \left( a^{\dagger }-a\right) \right] \left\vert n\right\rangle
\notag \\
&=&\frac{2\beta \eta }{n+1}e^{-2\beta ^{2}\eta ^{2}}L_{n}^{1}(4\beta
^{2}\eta ^{2}).
\end{eqnarray}%
By employing the similar approximation, we keep the leading terms by
expanding
\begin{eqnarray}
&&\left( a^{\dagger }-a\right) \cosh [2\beta \eta \left( a^{\dagger
}-a\right) ]  \notag \\
&=&T(a^{\dagger }a)a^{+}-aT(a^{\dagger }a)+O(\beta ^{3}\eta ^{3}),
\end{eqnarray}
and
\begin{eqnarray}
&&\left( a^{\dagger }-a\right) \sinh [2\beta \eta \left( a^{\dagger
}-a\right) ]\}  \notag \\
&=&D^{0}\left( a^{\dagger }a\right) +D^{2}\left( a^{\dagger }a\right)
a^{\dagger 2}+a^{2}D^{2}\left( a^{\dagger }a\right) +O(\beta ^{4}\eta ^{4}),
\notag \\
&&
\end{eqnarray}%
where the coefficients $T(a^{\dagger }a)$, $D^{0}\left( a^{\dagger }a\right)
$ and $D^{2}\left( a^{\dagger }a\right) $ are obtained as $T_{n+1,n}$, $%
D_{n,n}^{0}$ and $D_{n+2,n}^{2}$ in the oscillator basis $|n\rangle $
respectively (see Appendix A).

After such procedure, we obtain an effective Hamiltonian $H_{3}=H^{\texttt{GSRWA}}
+\tilde{H_{1}}+\tilde{H_{2}}$, consisting of
\begin{eqnarray}\label{HGSRWA}
H^{\texttt{GSRWA}} &=&\eta _{0}+\eta _{1}\omega a^{\dagger }a+\sigma _{z}[\frac{%
\Delta }{2}G_{0}(a^{\dagger }a)-\gamma \eta D_{0}(a^{+}a)]  \notag \\
&&+[(\alpha -\omega \beta )\eta _{3}+\frac{\Delta }{2}F\left( a^{\dagger
}a\right) -\gamma \eta T(a^{\dagger }a)]a^{\dagger }\sigma _{-}  \notag \\
&&+H.c.,
\end{eqnarray}
\begin{eqnarray}
\tilde{H_{1}} &=&[(\alpha -\omega \beta )\eta _{3}-\frac{\Delta }{2}F\left(
a^{\dagger }a\right) +\gamma \eta T(a^{\dagger }a)]a^{\dagger }\sigma
_{+}+H.c.,  \notag \\
&& \\
\tilde{H_{2}} &=&(a^{\dagger 2}+a^{2})\eta _{2}+\sigma _{z}\{\frac{\Delta }{2%
}[a^{2}G_{2}(a^{\dagger }a)+G_{2}(a^{\dagger }a)a^{\dagger 2}]  \notag \\
&&-\gamma \eta \lbrack a^{2}D_{2}(a^{+}a)+D_{2}(a^{+}a)a^{\dagger 2}]\}.
\end{eqnarray}
The transformed Hamiltonian $H^{\texttt{GSRWA}}$ includes the additional squeezing
transformation and retains the mathematical structure of the
ordinary RWA, so-called the generalized squeezing RWA (GSRWA) Hamiltonian.
And $\tilde{H_{1}} $ and $\tilde{H_{2}}$ represent the CRW coupling
and the two-excitation process.

We require that the CRW term $\tilde{H_{1}}$ and two-excitation
term $\tilde{H_{2}}$ vanish by choosing the form of
displacement $\beta $ and squeezing $\lambda $. Firstly, the matrix elements
$\langle n+1,+z|\tilde{H_{1}}|n,-z\rangle $ for the CRW terms equals to zero,
where $|\pm z\rangle $ denotes the eigenstates of $\sigma _{z}$. It
yields the equation%
\begin{equation}
(\alpha -\omega \beta )\eta _{3}-\frac{\Delta }{2}F_{n+1,n}+\gamma \eta
T_{n+1,n}=0.  \label{bt1}
\end{equation}%
Secondly, by projecting the two-excitation Hamiltonian to $\langle n+2|%
\tilde{H_{2}}|n\rangle $, one obtain
\begin{equation}
\eta _{2}-(\frac{\Delta }{2}G_{n+2,n}^{2}-\gamma \eta D_{n+2,n}^{2})=0.
\label{bt2}
\end{equation}%
The variational displacement $\beta $ and squeezing $\lambda $ is determined
by solving the Eqs.(\ref{bt1}) and (~\ref{bt2}) in detail in the Appendix B.
The analytical solutions of the squeezing $\lambda $ and displacement $\beta
$ are interesting since they play a crucial role in giving the explicit
energy spectrums and eigenfunctions. The nonlinear equations in Eqs.(~\ref%
{bt1p}) and (~\ref{bt2p}) cannot be solved analytically. When the
parameters $\lambda $ and $\beta $ is small compared with the unit,
the two nonlinear equations are simplified in the Appendix B,
resulting in analytical solutions
\begin{equation}  \label{lamb}
\lambda \simeq \frac{(\Delta \kappa ^{2}-2\gamma \kappa )(1-2\kappa ^{2})}{%
2\omega +4(\kappa ^{2}\Delta -2\gamma \kappa )(1-2\kappa ^{2})},
\end{equation}%
and%
\begin{equation}  \label{beta}
\beta^{\mathtt{GSRWA}} \simeq \frac{\alpha +\gamma e^{-4\lambda }e^{-2\kappa
^{2}\exp (-4\lambda )}}{\omega +\Delta e^{-4\lambda }e^{-2\kappa ^{2}\exp
(-4\lambda )}},
\end{equation}%
with $\kappa =(\alpha +\gamma)/(\omega +\Delta)$. On the other hand,
the GVM only with the displacement transformation $U$ is easily carried out
by setting the squeezing parameter $\lambda =0$ in Eq.(~\ref{bt1}),
resulting in the displacement $\beta^{\mathtt{GVM}} \simeq (\alpha +\gamma
e^{-2\kappa ^{2}})/(\omega +\Delta e^{-2\kappa ^{2}})$.

Consequencely, we present a solvable Hamiltonian $H^{\texttt{GSRWA}}$ (~\ref{HGSRWA})
by eliminating the CRT terms $\tilde{H_{1}}$ and two-excitation terms $\tilde{H_{2}}$.
The simplicity of the approximation is
based on its close connection to the standard RWA, giving analytical
eigenstates and eigenenergies. Our aim is to improve the
GVM with only the displacement tranformation to our GSRWA with the additional
squeezing transformation. Similar to the GVM employed in the isotropic Rabi model~\cite{yu},
one-excitation terms are kept as $F\left( a^{\dagger
}a\right)a^{\dagger }\sigma _{-}+H.C.$
And we extend the treatment to anisotropic Rabi case with additional terms
$T(a^{\dagger }a)a^{\dagger }\sigma _{-}+H.C$. Unlike the GVM, we take into account
the squeezing transformation and include the deformation effects of the oscillator state,
resulting in a displaced-squeezed oscillator state.
And the solvable Hamiltonian $H^{\mathtt{GSRWA}}$ involves
the effects of two-excitation process, which have completely ignored
in the GVM. Our approach is expected to
extend the range of validity to the npDSC regime through
involving effects of displacement and deformations.

\section{Energy spectrum}
Now we investigate the advantage of the GSRWA in terms of the excited states
and energy levels, revealing the failure of the GVM underestimated
the squeezing transformation in the npDSC regime.

One can easily diagonalize the Hamiltonian (~\ref{HGSRWA}) in the basis
of $|+,n\rangle $ and $|-,n+1\rangle $ ($n\geq 0$),
\begin{widetext}
\begin{equation}
H^{\mathtt{GSRWA}}=\left(
\begin{array}{ll}
\omega \eta _{1}n+\eta _{0}+f(n) & R_{n,n+1}\sqrt{n+1} \\
R_{n+1,n}\sqrt{n+1} & \omega \eta _{1}(n+1)+\eta _{0}-f(n+1)%
\end{array}%
\right) ,
\end{equation}
\end{widetext}with $f(n)=\frac{\Delta }{2}G_{n,n}^{0}-\gamma \eta
D_{n,n}^{0} $ and $R_{n+1,n}=(\alpha -\beta )\eta _{3}+\frac{\Delta }{2}%
F_{n+1,n}\left( n\right) -\gamma \eta T_{n+1,n}\left( n\right) $. The GSRWA
is identical in form to the corresponding term in the usual RWA Hamiltonian.
Solving the blocks of the GSRWA matrix form yields the eigenvalues
\begin{eqnarray}
E_{n}^{\pm} &=&(n+\frac{1}{2})\eta _{1}+\eta _{0}+\frac{1}{2}%
(R_{n,n}-R_{n+1,n+1})  \notag  \label{energy} \\
&&\pm \frac{1}{2}\sqrt{[\eta _{1}-(R_{n,n}+R_{n+1,n+1})]^{2}-4R_{n+1,n}^{2}},
\end{eqnarray}%
and the corresponding eigenfunctions
\begin{eqnarray}  \label{eigenstate0}
|\varphi _{+,n}\rangle &=&\cos \frac{\theta _{n}}{2}|n\rangle |+z\rangle
+\sin \frac{\theta _{n}}{2}|n+1\rangle |-z\rangle ,
\end{eqnarray}
\begin{eqnarray}  \label{eigenstate}
|\varphi _{-,n}\rangle &=&\sin \frac{\theta _{n}}{2}|n\rangle |+z\rangle
-\cos \frac{\theta _{n}}{2}|n+1\rangle |-z\rangle ,
\end{eqnarray}%
where $\theta _{n}=\arccos (\delta _{n}/\sqrt{\delta _{n}^{2}+4R_{n+1,n}^{2}}%
)$, and $\delta _{n}=-\omega \eta _{1}+f(n)+f(n+1)$.
For the original Hamiltonian $H$ in Eq.(~\ref{horig}) with CRW terms, eigenstates can be
obtained using the unitary transformations $U$ and $S$ in the following
\begin{eqnarray}  \label{eigenstate1}
|\Psi _{+,n}\rangle &=&U^{\dagger }S^{\dagger }|\varphi _{+,n}\rangle  \notag
\\
&=&\frac{1}{\sqrt{2}}[(\sin \frac{\theta _{n}}{2}|n+1\rangle _{+,ds}-\cos
\frac{\theta _{n}}{2}|n\rangle _{+,ds})|+x\rangle  \notag \\
&&+(\sin \frac{\theta _{n}}{2}|n+1\rangle _{-,ds}+\cos \frac{\theta _{n}}{2}%
|n\rangle _{-,ds})|-x\rangle ],  \notag \\
&&
\end{eqnarray}
\begin{eqnarray}  \label{eigenstate2}
|\Psi _{-,n}\rangle &=&U^{\dagger }S^{\dagger }|\varphi _{-,n}\rangle  \notag
\\
&=&\frac{1}{\sqrt{2}}[(-\cos \frac{\theta _{n}}{2}|n+1\rangle _{+,ds}-\sin
\frac{\theta _{n}}{2}|n\rangle _{+,ds})|+x\rangle  \notag \\
&&+(-\cos \frac{\theta _{n}}{2}|n+1\rangle _{-,ds}+\sin \frac{\theta _{n}}{2}%
|n\rangle _{-,ds})|-x\rangle ],  \notag \\
&&
\end{eqnarray}%
where $|\pm x\rangle =(\pm |+z\rangle +|-z\rangle )/\sqrt{2}$ is the
eigenstate of $\sigma _{x}$. And the displaced-squeezed oscillator state is
\begin{equation}
|n\rangle _{\pm ,ds}=e^{\mp \beta (a^{\dagger }-a)}e^{\lambda
(a^{2}-a^{\dagger 2})}|n\rangle, \label{dsstate}
\end{equation}%
which describes both the displacement and deformation effects of the oscillator
states induced by the atom-cavity coupling.

Meanwhile, under the GVM by only adjusting the displacement to eliminate the
CRW terms, the analytical eigenvalues $E_{\pm,n }^{\mathtt{GVM}}$ and
eigenstates $|\varphi _{\pm,n}^{\mathtt{GVM}}\rangle$ for the anisotropic Rabi model is obtained by setting
$\beta =\beta ^{\mathtt{GVM}}$ and $\lambda =0$ in Eqs.(~\ref{energy})-(~\ref%
{eigenstate}). The corresponding eigenstates for the original Hamiltonian in
the GVM can be derived using only the displacement transformations as $|\Psi
_{-,n}^{\mathtt{GVM}}\rangle=U^\dagger|\varphi _{\pm,n}^{\mathtt{GVM}%
}\rangle $, and the displaced-squeezed state $%
|n\rangle _{\pm ,ds}$ in Eqs.(~\ref{eigenstate1}) and (~\ref{eigenstate2})
is replaced by the displaced state
\begin{equation}
|n\rangle _{\pm ,d}=e^{\mp
\beta (a^{\dagger }-a)}|n\rangle.
\end{equation}
Due to the peculiarities associated with the displaced-squeezed state, we
examine the energy levels to test the accuracy of the GSRWA.

\begin{figure}[tbp]
\includegraphics[scale=0.75]{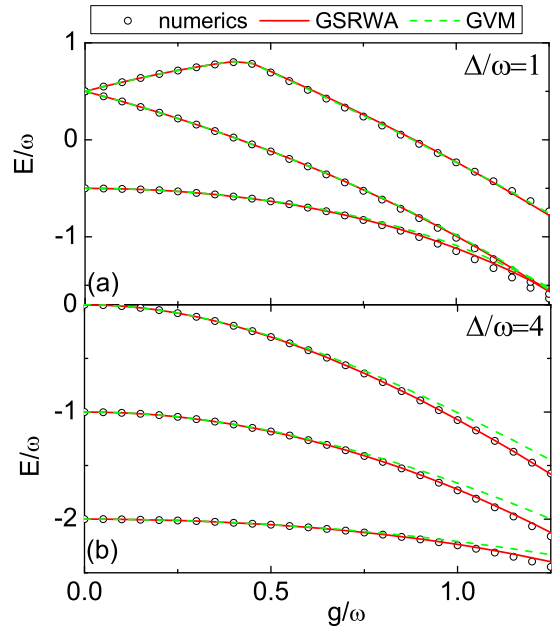}
\caption{(Color online) Energy levels $E_{n}/\protect\omega $ in the
isotropic case as a function of $g/\protect\omega $ for $\Delta /\protect%
\omega =1$(a) and $\Delta/\protect\omega=4$ (b) by means of the GSRWA (solid
lines), numerical simulation (circles), GVM (dash-dotted lines) and GRWA
(dashed lines)for the isotropic case $\tau=1$.}
\label{fig1}
\end{figure}
The energy levels from the numerical solution of the full Hamiltonian (~\ref{horig}),
the GVM, and the GSRWA are plotted for the isotropic case $\tau=1$
in Fig.~\ref{fig1}. The GSRWA with optimal displacement $\beta$ in Eq.(~\ref{beta})and
squeezing $\lambda$ in Eq.(~\ref{lamb}) captures the behavior of energy levels,
and provides an agreement with the numerical ones
ranging from the ultra-strong to npDSC
regimes. The GVM with only the displacement transformation
produces the correct behavior in the ultra-strong coupling regime, but
breaks down in the npDSC regime $g/\omega >0.7$.
The failure becomes more pronounced as the atom frequency $\Delta/\omega$ increases up
to $4$ in Fig.~\ref{fig1}(b), displaying a noticeable divergence of the GVM.
It reveals that the displaced state is not a reasonable treatment in the npDSC regime, where
the displaced-squeezed state is preferable and the
deformation effects is appreciable.

\begin{figure}[tbp]
\includegraphics[scale=0.75]{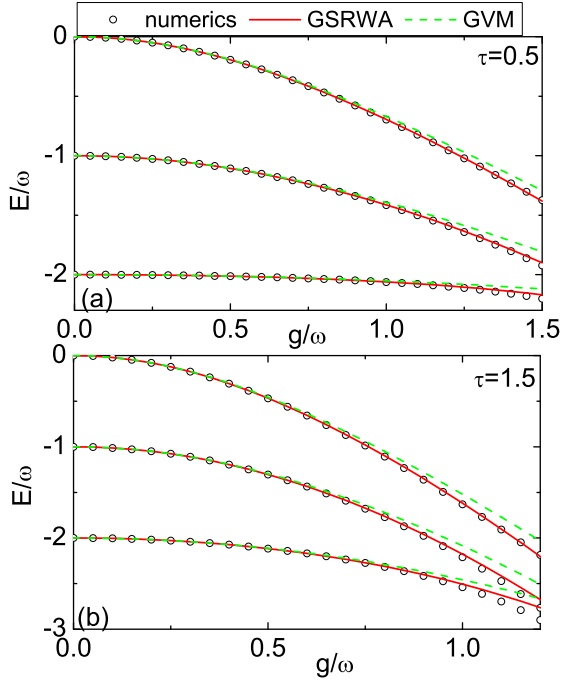}
\caption{(Color online) Energy-level crossing as a function of $g/\protect%
\omega$ in the anisotropic Rabi case (a) $\tau=0.5$ (b) $\tau=1.5$ for
the detuning parameter $\Delta/\protect\omega=4$ by means of the GSRWA
(solid lines), numerical simulation (circles), GVM (dash-dotted lines) for
the anisotropic case $\tau=0.5$.}
\label{fig2}
\end{figure}

Fig.~\ref{fig2} shows energy levels for the anisotropic Rabi case
with relative weight $\tau=1.5$ and $0.5$ for the high-frequency atom $\Delta/\omega=4$.
For small weight of the CRW interactions with $%
\tau=0.5$ in Fig.~\ref{fig2}(a), the GSRWA is surprising robust
as the coupling strength increases up to $%
g/\omega \sim 1.5$, where the energies in the GVM
show dramatic deviation. Moreover, the GVM gets worse as the relative weight of
the CRW terms increases to $\tau =1.5$ in Fig.~\ref%
{fig2}(b). It exhibits an overall improvement of the GSRWA
with the displaced-squeezed state to the GVM with the displaced state
as the relative weight between the rotating-wave and CRW interactions increases .
The advantage of our GSRWA lies in the contribution
from the squeezing and displacement of the oscillator state.
The GVM fails in particular to describe the eigenstates with the displaced state,
which should be more sensitive in characterize the squeezing effects
and the quantum dynamics presented in the following.

\section{Squeezing effects}

We analyze the displaced-squeezed state in the GSRWA to
explore the deformation or squeezing effects,
which are described by the quadrature variance for momentum operator in the ground state.
The ground state for the GSRWA is just as in the RWA giving by $%
|0\rangle |-z\rangle$. The operators
expectation values of the ground state follows that
\begin{eqnarray}
\langle a\rangle &=&\langle -z|\langle 0|SUaU^{\dagger }S^{\dagger }|0\rangle |-z\rangle
\notag \\
&=&\langle 0|a^{\dagger }\sinh 2\lambda +a\cosh 2\lambda |0\rangle -\beta
\notag \\
&=&-\beta ,
\end{eqnarray}%
and
\begin{eqnarray}
\langle a^{\dagger }a\rangle
&=&\langle -z|\langle 0|SUa^{\dagger }U^{\dagger }S^{\dagger }SUaU^{\dagger
}S^{\dagger }|0\rangle |-z\rangle  \notag \\
&=&\sinh ^{2}2\lambda +\beta ^{2}.
\end{eqnarray}
The variance $\Delta p$ of the momentum $p=i\sqrt{\frac{\omega }{2}}%
(a^{\dagger }-a)$ can be determined from these expectation values, so that
\begin{eqnarray}  \label{variancep}
\Delta p &=&\langle p^{2}\rangle -\langle p\rangle ^{2}  \notag \\
&=&-\frac{\omega }{2}\langle a^{\dagger 2}+a^{2}-2a^{\dagger }a-1\rangle
\notag \\
&=&\frac{\omega }{2}e^{-4\lambda }.
\end{eqnarray}
Similarly, the variance $\Delta x$ of the position $x=(a^{\dagger }+a)/\sqrt{2\omega}$
is given by $\Delta x=e^{4\lambda }/(2\omega)$.
The uncertainty in the momentum and position variables are therefore easily
obtained as $\Delta p\Delta x=1/2$, which satisfy the minimum-uncertainty
relation for the displaced-squeezed state.

\begin{figure}[tbp]
\includegraphics[scale=0.85]{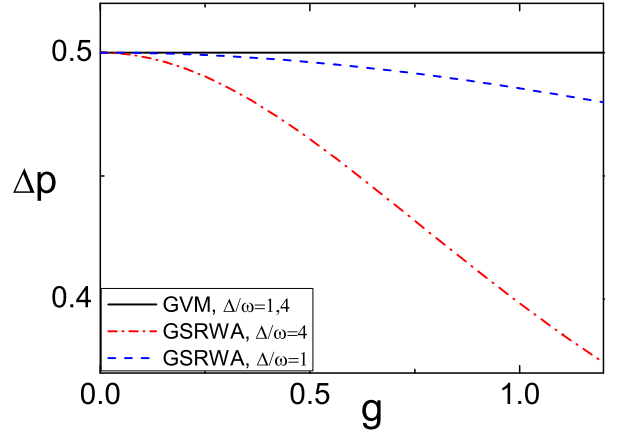}
\caption{(Color online) Squeezing effect with the momentum variance $\Delta p$
as a function of $g$ for different atom frequency $\Delta/\protect\omega%
=1$ (green dashed line) and $\Delta/\omega=4$ (red dashed-dotted line) obtained by the GSRWA
and the GVM (black solid line) in the isotropic case.}
\label{fig3}
\end{figure}

Meanwhile, the variance of momentum $\Delta p$ in
the GVM equals to $0.5$, which can be obtained easily from Eq.(~\ref
{variancep}) with the displaced state. Fig.~\ref{fig3} displays that
the momentum variance by the GSRWA is smaller than $0.5$, indicating that the
momentum quadrature is squeezed with the displaced-squeezed state.
The quantum fluctuations in momentum variable are reduced
at the expense of the
corresponding increased fluctuations in the position variable such that the
uncertainty relation is not violate. The
squeezing effect is accurately captured by the displaced-squeezed state.

\section{Population dynamics}
The dynamical behavior of the two-level atom is of particular interest. In
this section we explore the atomic population dynamics in the
anisotropic Rabi model to test the accuracy of the energies and
eigenstates in the npDSC regimes.

The initial state is taken to be $|\varphi (0)\rangle =|-x\rangle |\alpha
_{-1}\rangle $ with the coherent state for the oscillator $|\alpha
_{-1}\rangle =e^{\beta (a^{\dagger }-a)}|\alpha \rangle $. The wave function
evolutes as $|\varphi (t)\rangle =e^{-iHt}|\varphi (0)\rangle $, which can
be expanded by the eigenvalues $\{E_{n}\}$ (~\ref{energy}) and eigenstates $%
\{|\Psi _{\pm ,n}\rangle \}$ in Eqs.(~\ref{eigenstate1}) and (~\ref%
{eigenstate2}) in the GSRWA.

The population for the atom remaining in the initial state $|-x\rangle $ is
given by $P_{-1}(t)=|\langle -x|\mathtt{Tr}_{\mathtt{ph}}|\varphi (t)\rangle
\langle \varphi (t)|-x\rangle |$, which is derived explicitly in the
Appendix C. From the population formula in Eq.(~\ref{population}), function $%
S_{n}$ displays the frequency of the Rabi's oscillation depending on the
transition frequencies $\Delta E_{m,n}^{j,j^{\prime }}=E_{j,m}-E_{j^{\prime
},n}$ with $m=n,n-1$ ($j,j^{\prime}=\pm$).

\begin{figure}[tbp]
\includegraphics[scale=0.75]{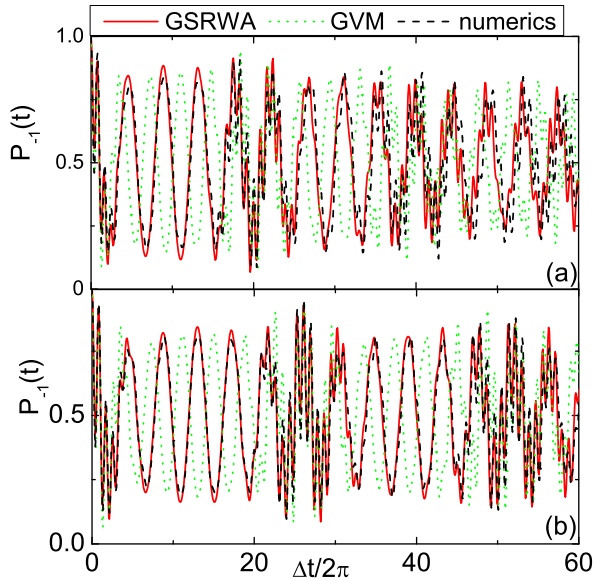}
\caption{(Color online) Population $P_{-1}(t)$ for the coupling strength $%
g/\protect\omega=0.5$ for the isotropic case $\tau=1$(a) and anisotropic
case $\tau=0.5$ (b) by means of the GSRWA (solid lines), numerical
simulation (circles), GVM (dash-dotted lines) and GRWA (dashed lines).}
\label{fig4}
\end{figure}

Figure ~\ref{fig4} shows the population $P_{-1}(t)$ as a function of the
scaled time $\Delta t/2\pi$ at npDSC strength $g/\omega=0.5$ for high-frequency
atom $\Delta/\omega=4$. We compare exact numerical results to
the GSRWA and the GVM. Obviously, qualitative agreement between the GSRWA
and the numerical ones of the dynamics oscillation is quite good even for
long time scale for the isotropic and anisotropic Rabi model. However,
the results in the GVM are quite different
from the numerical ones. Apart from the energy levels, also the eigenstates
become now of importance. The failure of population dynamics by the GVM is
due to the breaks down of the displaced state in the npDSC regime, where the
displaced-squeezed state is more stable to capture dynamics.

\section{conclusion}

We study the anisotropic Rabi model analytically
in the nonperturbative DSC regime, belonging to the region between the ultra-strong
and perturbative DSC coupling regimes.
The GSRWA is performed by adding a squeezing transformation
to the existing solutions with only the displacement transformation,
giving an solvable Hamiltonian in the same form of the standard RWA.
Energy levels obtained by the GSRWA agree well with numerical
ones in a wide range of coupling strength, whereas the previous results show distinguished
deviation in the nonperturbative DSC. Due to the displaced-squeezed state, the squeezed momentum
variance displays the deformation effects induced by the atom-cavity coupling,
which is omitted in the previous methods with the displaced state.
And the population dynamics by the GSRWA is robust in the
nonperturbative DSC regime even for high-frequency atom.
The advantage of our GSRWA is not only substantial
improvement of energy levels but also the stability of the
displaced-squeezed oscillator state. Our approach provides
an easy-to-implement analytical solutions to qubit-oscillator
coupling systems currently for ultra-strong and perturbative DSC strengths,
and also motivates further studies of multi-modes spin-boson model.

\acknowledgments
This work was supported by the Chongqing Research Program of Basic Research
and Frontier Technology (Grant No.cstc2015jcyjA00043), and the Research Fund
for the Central Universities (Grants No.106112016CDJXY300005, and No.
CQDXWL-2014-Z006).

$^{*}$ Email:yuyuzh@cqu.edu.cn

\appendix

\section{Expanding of even and odd function}

Since $\cosh \left[ 2\beta \eta \left( a^{\dagger }-a\right) \right] $ is
expanded as $G^{0}\left( a^{\dagger }a\right) +G^{2}\left( a^{\dagger
}a\right) a^{\dagger 2}+a^{2}G^{2}\left( a^{\dagger }a\right) +O(\beta
^{4}\eta ^{4})$, coefficient $G_{2}\left( a^{\dagger }a\right) $ of the
two-excitation terms can be derived in the oscillator basis $|n\rangle $ as
\begin{eqnarray}
G_{n+2,n}^{2} &=&\frac{1}{\sqrt{(n+2)(n+1)}}\langle n+2|\cosh [2\beta \eta
\left( a^{\dagger }-a\right) ]|n\rangle  \notag \\
&=&\frac{4\beta ^{2}\eta ^{2}}{(n+2)(n+1)}e^{-2\beta ^{2}\eta
^{2}}L_{n}^{2}(4\beta ^{2}\eta ^{2}).
\end{eqnarray}%
Similarily, coefficients $T(a^{\dagger }a)$, $D_{0}\left( a^{\dagger
}a\right) $ and $D_{2}\left( a^{\dagger }a\right) $ in the odd function $%
\left( a^{\dagger }-a\right) \cosh [2\beta \eta \left( a^{\dagger }-a\right)
]$ and even function $\left( a^{\dagger }-a\right) \sinh [2\beta \eta \left(
a^{\dagger }-a\right) ]$ are given as $T_{n+1,n}$, $D_{n,n}^{0}$ and $%
D_{n+2,n}^{2}$ respectively

\begin{eqnarray}
T_{n+1,n} &=&\frac{1}{\sqrt{n+1}}\left\langle n+1\right\vert \left(
a^{\dagger }-a\right) \cosh [2\beta \eta \left( a^{\dagger }-a\right)
]|n\rangle  \notag \\
&=&G_{n,n}^{0}-(n+2)G_{n+2,n}^{2},
\end{eqnarray}%
\begin{eqnarray}
D_{n,n}^{0} &=&\langle n|\left( a^{\dagger }-a\right) \sinh [2\beta \eta
\left( a^{\dagger }-a\right) ]|n\rangle  \notag \\
&=&-\sqrt{n}F_{n-1,n}(n)-\sqrt{n+1}F_{n+1,n}(n),
\end{eqnarray}

and%
\begin{eqnarray}
D_{n+2,n}^{2} &=&\frac{\langle n+2|\left( a^{\dagger }-a\right) \sinh
[2\beta \eta \left( a^{\dagger }-a\right) ]|n\rangle }{\sqrt{(n+1)(n+2)}}
\notag \\
&=&F_{n+1,n}(n)-\frac{\sqrt{n+3}}{\sqrt{(n+1)(n+2)}}F_{n+3,n}(n),  \notag \\
&&
\end{eqnarray}

with $F_{n+3,n}(n)=\langle n+3|\sinh [2\beta \eta \left( a^{\dagger
}-a\right) ]|n\rangle =(2\beta \eta )^{3}e^{-2\beta ^{2}\eta
^{2}}L_{n}^{3}(4\beta ^{2}\eta ^{2})/\sqrt{(n+1)(n+2)(n+3)}$.

\section{Solution equations of $\lambda $ and displacement $\beta$}

To obtain the optimal squeezing parameter $\lambda $ and displacement $\beta
$ from Eqs.(~\ref{bt1}) and (~\ref{bt2}), it is equivalent to solve the
equations in detail
\begin{eqnarray}
0 &=&(\alpha -\omega \beta )\eta _{3}-\frac{\Delta \beta \eta }{n+1}%
e^{-2\beta ^{2}\eta ^{2}}L_{n}^{1}(4\beta ^{2}\eta ^{2})  \notag
\label{bt1p} \\
&&+\gamma \eta e^{-2\beta ^{2}\eta ^{2}}[L_{n}(4\beta ^{2}\eta ^{2})-\frac{%
4\beta ^{2}\eta ^{2}}{n+1}L_{n}^{2}(4\beta ^{2}\eta ^{2})],
\end{eqnarray}

\begin{eqnarray}
0 &=&\eta _{2}-\frac{2\Delta \beta ^{2}\eta ^{2}}{(n+2)(n+1)}e^{-2\beta
^{2}\eta ^{2}}L_{n}^{2}(4\beta ^{2}\eta ^{2})  \notag  \label{bt2p} \\
&&+\gamma \eta e^{-2\beta ^{2}\eta ^{2}}[\frac{2\beta \eta }{n+1}%
L_{n}^{1}(4\beta ^{2}\eta ^{2})  \notag \\
&&-\frac{(2\beta \eta )^{3}}{(n+1)(n+2)}e^{-2\beta ^{2}\eta
^{2}}L_{n}^{3}(4\beta ^{2}\eta ^{2})].
\end{eqnarray}
When the parameters $\lambda $ and $\beta $ are small compared with the unit,
the associated Lagurre polynomial is given approximately by $L_{n}^{1}(4\beta
^{2}\eta ^{2})\simeq n+1,$ $L_{n}^{2}(4\beta ^{2}\eta ^{2})\simeq
(n+1)(n+2)/2$ and $L_{n}^{3}(4\beta ^{2}\eta ^{2})\simeq (n+1)(n+2)(n+3)/3!$.
Thus the above nolinear equations are simplified as
\begin{eqnarray}
0 &=&(\alpha -\omega \beta )-\Delta \beta e^{-4\lambda }e^{-2\beta ^{2}\eta
^{2}}  \notag  \label{bp11} \\
&&+\gamma e^{-4\lambda }[1-2(n+2)\beta ^{2}\eta ^{2}]e^{-2\beta ^{2}\eta
^{2}},
\end{eqnarray}%
and
\begin{eqnarray}
0 &=&(e^{4\lambda }-e^{-4\lambda })-4\Delta \beta ^{2}e^{-4\lambda
}e^{-2\beta ^{2}\eta ^{2}}  \notag  \label{bp22} \\
&&+4\gamma e^{-4\lambda }[2\beta -\frac{4}{3}(n+4)\beta ^{3}e^{-4\lambda
}]e^{-2\beta ^{2}\eta ^{2}}.
\end{eqnarray}

\section{Analytical expression of population}
The wave function $|\varphi (t)\rangle $ can be expanded by the eigenvalues $%
\{E_{n}\}$ and eigenstates $\{|\Psi _{\pm ,n}\rangle \}$ as
\begin{equation}
|\varphi (t)\rangle =f_{0}e^{-iE_{0}t}|\Psi _{0}\rangle +\sum_{n=0}f_{\pm
,n}e^{-iE_{\pm ,n}t}|\Psi _{\pm ,n}\rangle ,
\end{equation}%
where the coefficients $f_{0}=\langle \Psi _{0}|\varphi (0)\rangle $ and $%
f_{\pm ,n}=\langle \Psi _{\pm ,n}|\varphi (0)\rangle $. And the overlap
between the displaced-squeezed state and the initial coherent state is
expressed by the polynomials
\begin{eqnarray}
&&_{-,ds}\langle n|\alpha _{-1}\rangle  \notag \\
&=&\chi \sum_{i=0}^{n/2}\frac{(-0.5\tanh 2\lambda )^{i}}{i!(n-2i)!}%
e^{(n-2i+1/2)In\sec h2\lambda }(-\alpha )^{n-2i}  \notag \\
&&
\end{eqnarray}%
with $\mu =e^{2\lambda }$, $\sec h(2\lambda )=\frac{2\mu }{1+\mu ^{2}}$and $%
\tanh (2\lambda )=\frac{\mu ^{2}-1}{1+\mu ^{2}}$, and $\chi =\sqrt{n!}%
e^{-\alpha ^{2}/2}e^{\frac{\alpha ^{2}}{2}\tanh 2\lambda }$. Thus, the
coefficient $C_{-x}$ of the atom state $|-x\rangle $ is

\begin{eqnarray}
C_{-x} &=&(\kappa _{0}e^{-iE_{0}t}+\kappa _{+,0}e^{-iE_{+,0}t}+\kappa
_{-,0}e^{-iE_{-,0}t})|0\rangle _{-,ds}  \notag \\
&&+\sum_{n>0,j=\pm }(\kappa _{j,n-1}e^{-iE_{j,n-1}t}+\kappa
_{j,n}e^{-iE_{j,n}t})|n\rangle _{-,ds},  \notag \\
\end{eqnarray}%
where coefficients are given as $\kappa _{0}=f_{0}/\sqrt{2}$, $\kappa
_{+,n}=\cos \frac{\theta _{n}}{2}f_{+,n}/\sqrt{2}$ , $\kappa _{-,n-1}=-\cos
\frac{\theta _{n-1}}{2}f_{-,n}/\sqrt{2}$, $\kappa _{+,n-1}=\sin \frac{\theta
_{n-1}}{2}f_{+,n}/\sqrt{2}$ and $\kappa _{-,n}=\sin \frac{\theta _{n}}{2}%
f_{-,n}/\sqrt{2}$. The population $P_{-1}(t)=|C_{-x}^{\ast }C_{-x}|$ for the
atom remaining in the initial state $|-x\rangle $ is expressed as
\begin{eqnarray}
P_{-1}(t) &=&\kappa _{0}\kappa _{+,0}\cos [(E_{0}-E_{+,0})t]  \notag
\label{population} \\
&&+\kappa _{0}\kappa _{-,0}\cos [(E_{0}-E_{-,0})t]  \notag \\
&&+\kappa _{+,0}\kappa _{-,0}\cos [(E_{-,0}-E_{+,0})t] \\
&&+\sum_{n>0}S_{n}(t)+k,
\end{eqnarray}%
where
\begin{eqnarray}
S_{n}(t) &=&\sum_{j,j^{\prime }=\pm }\kappa _{j,n}\kappa _{j^{\prime
},n}\cos \Delta E_{n,n}^{j,j^{\prime }}+\kappa _{j,n-1}\kappa _{j^{\prime
},n}\cos \Delta E_{n-1,n}^{j,j^{\prime }}  \notag  \label{sn} \\
&&+\kappa _{j,n-1}\kappa _{j^{\prime },n-1}\cos \Delta
E_{n-1,n-1}^{j,j^{\prime }}
\end{eqnarray}%
with $\Delta E_{m,n}^{j,j^{\prime }}=E_{j,m}-E_{j^{\prime },n}$, ($m=n,n-1$)
and the constant $k=\kappa _{0}^{2}+\kappa _{+,0}^{2}+\kappa _{-,0}^{2}$

\end{document}